\documentclass[twocolumn,english,prb,superscriptaddress, twocolumn, nofootinbib]{revtex4-1}
\usepackage[T1]{fontenc}
\usepackage[latin9]{inputenc}
\usepackage{babel}
\usepackage{array}
\usepackage{units}
\usepackage{multirow}
\usepackage{amsmath}
\usepackage{amssymb}
\usepackage{graphicx}
\usepackage{esint}
\usepackage[unicode=true,pdfusetitle,
 bookmarks=false,
 breaklinks=false,pdfborder={0 0 0},backref=false,colorlinks=false]
 {hyperref}
\usepackage{breakurl}

\makeatletter

\providecommand{\tabularnewline}{\\}

\usepackage{babel}
\usepackage{dcolumn}

\makeatother

\begin{document}

\title{Orbitally-driven insulator-metal transition in CuIr$_{2}$S$_{4}$:
Temperature dependent transient reflectivity study }

\author{M. Naseska}

\address{Complex Matter Department, Jozef Stefan Institute, Jamova 39, 1000
Ljubljana, Slovenia}

\author{P. Sutar}

\address{Complex Matter Department, Jozef Stefan Institute, Jamova 39, 1000
Ljubljana, Slovenia}

\author{D. Vengust}

\address{Complex Matter Department, Jozef Stefan Institute, Jamova 39, 1000
Ljubljana, Slovenia}

\author{S. Tsuchiya}

\address{Complex Matter Department, Jozef Stefan Institute, Jamova 39, 1000
Ljubljana, Slovenia}

\address{Department of Applied Physics, Hokkaido University, Sapporo, Japan}

\author{M. \v{C}eh}

\address{Centre for Electron Microscopy, Jozef Stefan Institute, Jamova 39,
1000 Ljubljana, Slovenia}

\author{D. Mihailovic}

\address{Complex Matter Department, Jozef Stefan Institute, Jamova 39, 1000
Ljubljana, Slovenia}

\address{Center of Excellence on Nanoscience and Nanotechnology Nanocenter
(CENN Nanocenter), Jamova 39, 1000 Ljubljana, Slovenia}

\author{T. Mertelj}

\email{tomaz.mertelj@ijs.si}

\selectlanguage{english}%

\address{Complex Matter Department, Jozef Stefan Institute, Jamova 39, 1000
Ljubljana, Slovenia}

\address{Center of Excellence on Nanoscience and Nanotechnology Nanocenter
(CENN Nanocenter), Jamova 39, 1000 Ljubljana, Slovenia}

\date{\today}
\begin{abstract}
Ultrafast transient reflectivity across the unusual three-dimensional
Peierls-like insulator-metal (IM) transition in CuIr$_{2}$S$_{4}$
was measured as a function of temperature. The low-temperature insulating-phase
transient response is dominated by broken-symmetry-induced coherent
lattice oscillations that abruptly vanish at the IM transition. The
coherent mode spectra are consistent with Raman spectra reported in
literature. The origin of the broken-symmetry-induced is also briefly
discussed.
\end{abstract}
\maketitle

\section{Introduction}

The spinel-structure compounds CuIr$_{2}$S$_{4}$ and MgTi$_{2}$O$_{4}$\citep{nagataHagino1994,IsobeUeda2002}
show unusual, presumably orbitally-driven,\citep{khomskiiMizokawa2005}
three-dimensional Peierls-like metal-insulator transitions on cooling
at $T_{\mathrm{IM}}\sim230$ K and $\sim260$K, respectively. The
transitions are accompanied with a long range charge ordering forming
Ir (Ti) dimers\citep{radaelliHoribe2002,schmidtRatcliff2004} and
a strong decrease of the magnetic susceptibility\citep{furubayashiMatsumoto1994,IsobeUeda2002}.

The lattice dynamical aspects of the the phase transition in CuIr$_{2}$S$_{4}$
were studied by means of the Raman spectroscopy\citep{zhangLing2010,zhangLing2013orbitally,MatsubaraEbihara2009}
and transient reflectivity\citep{MatsubaraEbihara2009}. It has been
suggested\citep{zhangLing2013orbitally} that the phonons associated
with the S-S bonds play a role in the unusual IM phase transition
so a more detailed insight into the lattice dynamical properties could
be beneficial for understanding the mechanism underlying the transition.

In Raman spectra several additional low frequency modes\citep{zhangLing2010}
appear in the low-$T$ insulating state in addition to the four characteristic
spinel-structure modes observed in the high-$T$ cubic metallic state.
The frequencies of these modes are, however, not consistent with the
coherent transient-reflectivity modes observed in an earlier transient
reflectivity study\citep{MatsubaraEbihara2009} in neither of the
phases. 

The difference might be related to the low temperature X-ray- and
visible-light-induced conducting phase\citep{ishibashiKoo2002,furubayashiSuzuki2003,takuboHirata2005,kiryukhinHoribe2006,takuboMizokawa2008}
with persistent dimerization\citep{ishibashiKoo2002,bozinMasadeh2011}
and suppressed long range order due to different optical excitation
in Raman and transient reflectivity experiments. To resolve the discrepancy
and get better insight into the low frequency lattice dynamics across
the insulator-metal (IM) transition in CuIr$_{2}$S$_{4}$ we performed
a detailed temperature dependent transient reflectivity study combined
with room-$T$ Raman spectroscopy characterization.

We find that the transient reflectivity at 1.55-eV photon energy in
the low-$T$ insulating (I) phase is dominated by the coherent lattice
response that abruptly vanishes at the IM transition. The frequencies
of the observed coherent modes are different than in the previous
transient reflectivity study\citep{MatsubaraEbihara2009} and are
consistent with the frequencies of the modes observed\citep{zhangLing2010}
by the Raman spectroscopy.

\section{Experimental}

\begin{figure}[b]
\includegraphics[width=0.9\columnwidth]{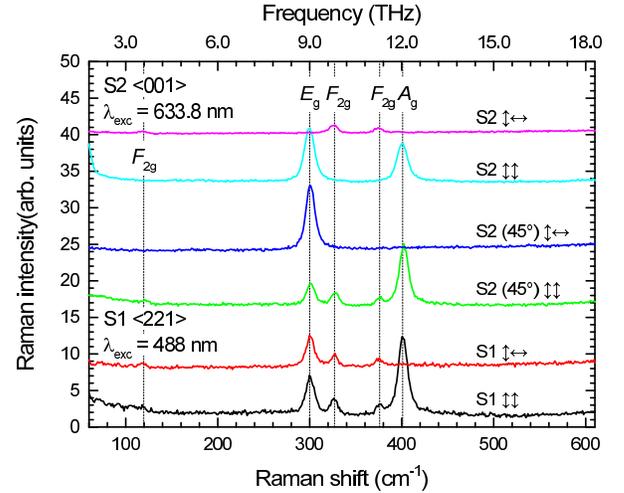}\caption{Room-$T$ micro-Raman spectra from cleaved crystal surfaces taken
in the back-scattering geometry. For sample S1 the in-plane excitation-light
polarization\label{fig:Raman} was random while for sample S2 the
polarization was either along {[}100{]} or {[}110{]}. The only $A_{\mathrm{g}}$
mode\citep{zhangLing2010} at 403 cm$^{-1}$ is always absent in the
crossed polarization spectra at both surface orientations while the
the $E_{g}$ mode at 302 cm$^{-1}$ and the two $F_{2g}$ modes at
327 and 375 cm$^{-1}$ are extinct only for polarizations along certain
high symmetry directions on the sample-S2 {[}001{]} surface. The traces
are vertically offset for clarity.}
\end{figure}

\begin{figure*}
\includegraphics[width=0.9\textwidth]{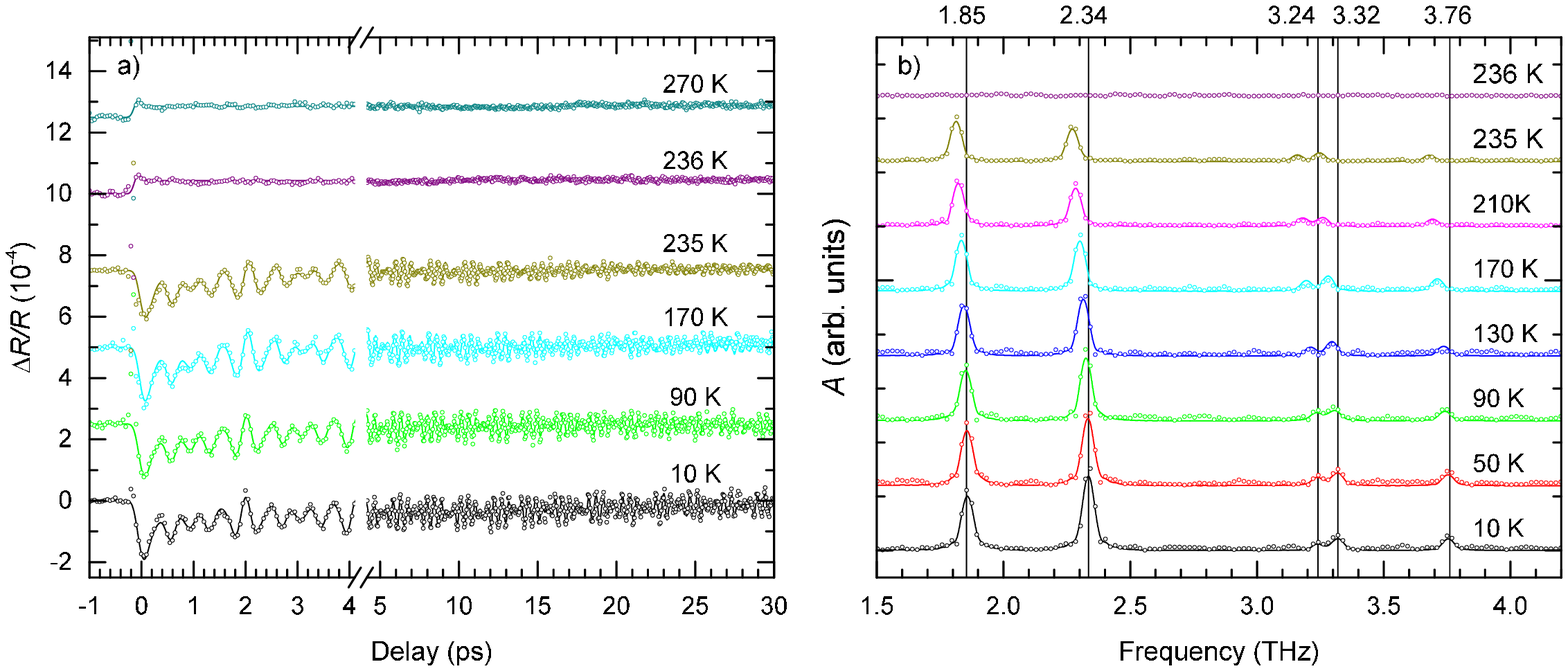}

\caption{a) The low-excitation-fluence transient reflectivity at selected temperatures
during the warming cycle in sample S1. Some coherent artifacts due
to the pump scattering are observed around zero delay. b) Fourier
transforms of the transient reflectivity at selected temperatures
during the warming cycle. The thin lines correspond to the DECP fits
discussed in text. Traces at different $T$ are vertically offset
for clarity.}
\label{fig:coherent}
\end{figure*}

\subsection{Methods and sample characterization}

Single crystals of CuIr$_{2}$S$_{4}$ were grown from Bi solution
as described in Ref. {[}\onlinecite{matsumotoNagata2000}{]}. The
crystal structure of the obtained millimeter-size crystals was checked
by means of X-ray diffraction. The quality of the crystals was evaluated
also by means of Raman scattering using 488-nm and 632.8-nm laser
excitation. Room-$T$ Raman spectra from as-grown crystal facets showed
four phonon lines at frequencies consistent with previous results\citep{zhangLing2010},
but somewhat broader. Polishing resulted in even broader Raman lines.
Cleaved surfaces, on the other hand, showed significantly narrower
and more intense Raman phonon lines (see Fig. \ref{fig:Raman}) with
an additional weak line at 120 cm$^{-1}$ and were used for all further
optical measurements. 

Here we present data from two cleaved crystals designated S1 and S2.
The orientation of the S1 cleaved surface was determined \emph{a posteriori}
from EBSD Kikuchi patterns to be close to the $\left\langle 221\right\rangle $
plane while the orientation of the S2 cleaved surface was inferred
from the Raman selection rules to be close to the $\left\langle 001\right\rangle $
plane.

The transient reflectivity measurements were performed using a standard
pump-probe setup using 50-fs linearly polarized laser pulses at 800
nm wavelength and the 250 kHz repetition rate as presented in detail
elsewhere.\citep{Mertelj2017} Both the pump- and probe-photon energy
were at the laser fundamental, $\hbar\omega=1.55$~eV. The transient
reflectivity, $\Delta R/R$, was measured at a near-normal incidence
from the cleaved surfaces. 

The pump and probe beam diameters were 40-70 and 18-30 $\mu$m, respectively.
The probe fluence was $\sim5$~$\mu$J/cm$^{2}$ for all measurements.
The Polarizations of pump and probe beams were perpendicular to each
other with a random orientation with respect to the crystal axes.
The transient reflectivity showed no polarization dependence.

\subsection{Results}

In Fig. \ref{fig:coherent} (a) we show the transient reflectivity
in sample S1\footnote{Both samples show identical $T$-dependence. }
at a few characteristic $T$ measured at weak excitation\footnote{The linearity of the response at this fluence was checked at 6 K,
190 K, 245 K and 270 K.} of $F=15$0 $\mu$J/cm$^{2}$ on a warming run. In the low-$T$ insulating
phase we observe a clear weakly damped coherent-phonon response that
is only slightly $T$-dependent. The beating pattern indicates the
presence of several modes. Fourier transform spectra {[}Fig. \ref{fig:coherent}
b){]} reveal two stronger modes at 1.85 and 2.34 THz together with
three weaker modes at 3.24, 3.32 and 3.76 THz. All the modes soften
and broaden with increasing $T$ and suddenly disappear at $T_{\mathrm{IM}\uparrow}=235$
K while on cooling they suddenly appear at $T_{\mathrm{IM}\downarrow}=233$
K. In the metallic phase above $T_{\mathrm{IM}}$\footnote{Due to the presence of small hysteresis we use $T_{\mathrm{IM}\uparrow}$
or $T_{\mathrm{IM}\downarrow}$ when we refer to the warming or cooling
runs, respectively, and $T_{\mathrm{IM}}$ when we do not care about
the the temperature change direction.} we observe no coherent modes and no relaxation\footnote{The traces show a hint of a weak sub-ps relaxation component that
is obscured by the coherent artifact.} on the 30-ps time scale of the experiment.

\section{Analysis}

\begin{figure}
\includegraphics[width=0.8\columnwidth]{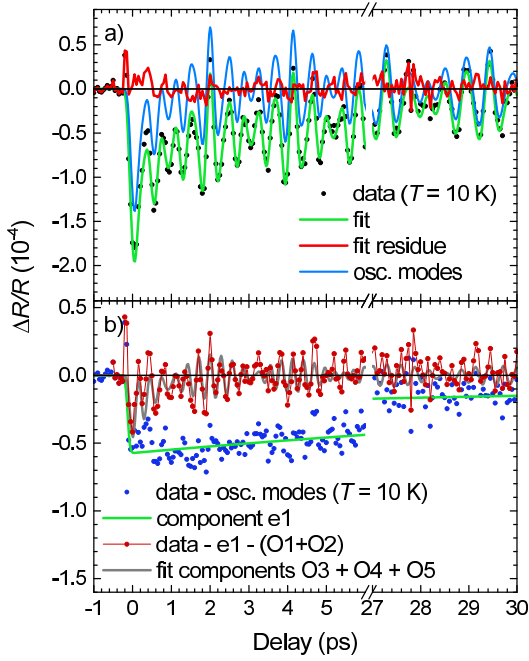}\caption{Illustration of different DECP model components in sample S1 at $T=10$
K. a) The total oscillatory part of the fit in comparison to the data,
fit and fit residue. b) Subtraction of various model component combinations
from the data with the corresponding remaining fit components. \label{fig:figFits}}
\end{figure}
\begin{figure}
\includegraphics[width=0.8\columnwidth]{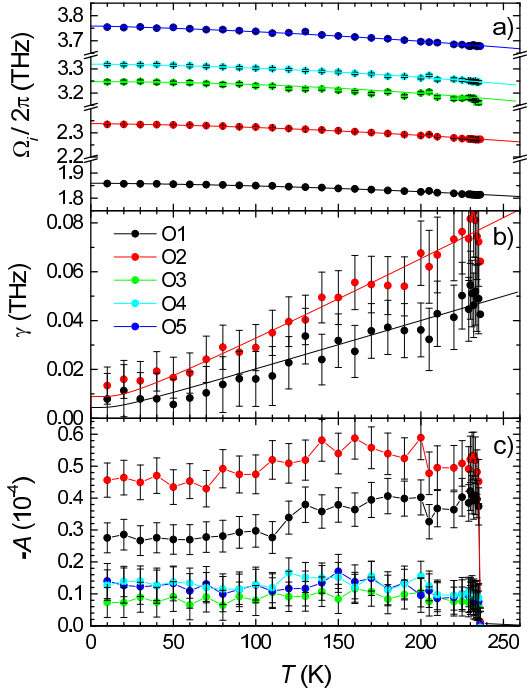}\caption{Temperature dependence of the coherent phonon parameters from DECP
fits on warming. The thin lines in a) are the anharmonic theory fits
of Eq. (3.9) from\citep{BalkanskiWallis1983} and in b) fits of Eq.
(3.8) from\citep{BalkanskiWallis1983}.}
\label{fig:OvsT}
\end{figure}
\begin{figure}
\includegraphics[width=0.8\columnwidth]{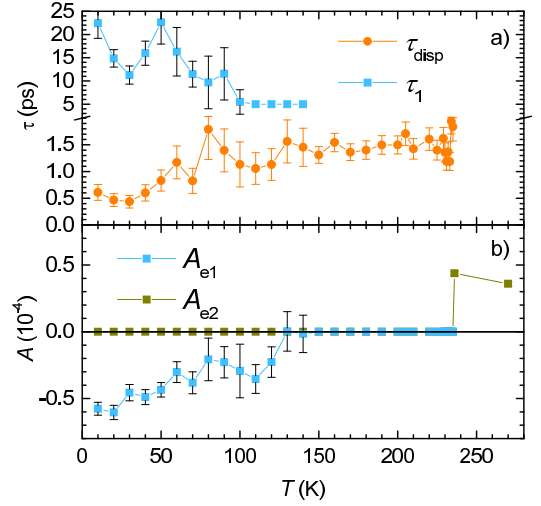}

\caption{Temperature dependence of the exponential fit-components parameters
at low excitation fluence from DECP fits on on warming.}
\label{fig:AvsT}
\end{figure}

\subsection{Data fitting}

The transient reflectivity data were analysed in the framework of
the displacive excitation of coherent phonons (DECP) theory\citep{zeiger1992theory}
(Appendix). The experimental transient reflectivity in the low-$T$
insulating phase can be fairly described assuming an exponentially
relaxing electronic displacive component with the relaxation time
$\tau_{\mathrm{displ}}$\footnote{Please see Appendix for the detailed definition.}
in the picosecond range that drives five oscillatory lattice modes
with the frequencies, $\Omega_{i}$, dampings, $\gamma{}_{i}$, and
amplitudes, $A_{\mathrm{O}i}$. An additional exponential component
needs to be included to completely describe the relaxation at low
$T$ (see Fig. \ref{fig:figFits}). 

Due to the small amplitudes of the three weakest coherent modes (O3,
O4, O5) the corresponding dampings showed large scattering and caused
fit instability. In order to prevent the adverse influence to the
fit convergence we fixed the three weakest modes $\gamma_{i}$ to
0.03 THz. The fixed values of the dampings were chosen to approximately
correspond to the values determined for the strongest modes O1 and
O2 in the middle of the relevant $T$ range. Since the scan lengths
are comparable or shorter than $\gamma_{i}^{-1}$ this introduces
only some systematic bias to the amplitudes of these modes, but does
not affect significantly their frequencies and the other-components
fit parameters.

It turns out that we can fit the data setting the displacive component
amplitude, $A_{\mathrm{displ}}=0$, in the full $T$-range below $T_{\mathrm{IM}}$
indicating that the displacive component only weakly couples to the
transient reflectivity at $\hbar\omega=1.55$-eV. Below $\sim130$~K,
however, inclusion of an additional exponentially relaxing component
with $\tau_{1}\sim15$ ps and amplitude $A_{\mathrm{e}1}$ was necessary
to completely fit the transient response. 

In Figs. \ref{fig:OvsT} and \ref{fig:AvsT} we show the temperature
dependence of the DECP theory parameters. The coherent modes amplitudes
show no or a rather weak $T$ dependence in the low-$T$ insulating
phase dropping abruptly to zero above $T_{\mathrm{IM}}$. With increasing
$T$ the modes show softening of about $\sim2\%$ accompanied by a
more significant damping increase. Both, the softening and the damping
increase can be attributed to the lattice anharmonicity.\citep{BalkanskiWallis1983}

Above $T_{\mathrm{IM}}$ the data can be fit by a double exponential
response with the relaxation times $\tau\sim100$~fs and $\tau_{\mathrm{e2}}\gg30$~ps.\footnote{The long lived component shows no relaxation in the experimental time
window.} Below $T_{\mathrm{IM}}$ both components abruptly vanish.

\subsection{Discussion}

\begin{table}
\begin{tabular}{c|c|r@{\extracolsep{0pt}.}lr@{\extracolsep{0pt}.}l}
Raman\footnote{Zhang \emph{et al., }Ref. {[}\onlinecite{zhangLing2010}{]}} & $\Delta R/R$\footnote{Matsubara \emph{et al.,} Ref. {[}\onlinecite{MatsubaraEbihara2009}{]}} & \multicolumn{4}{c}{$\Delta R/R$\footnote{present work}}\tabularnewline
($T=83$ K) & ($T=150$ K) & \multicolumn{4}{c}{($T=10$ K)}\tabularnewline
(cm$^{-1}$)  & (cm$^{-1}$)  & \multicolumn{2}{c}{(cm$^{-1}$) } & \multicolumn{2}{c}{(THz)}\tabularnewline
\hline 
?\footnote{Out of the measurement range in Ref. {[}\onlinecite{zhangLing2010}{]}.\label{fn:Out-of-range.}} & - & \multicolumn{2}{c}{62 } & 1&86\tabularnewline
?$^\mathrm{\ref{fn:Out-of-range.}}$ & - & \multicolumn{2}{c}{78} & 2&34\tabularnewline
?$^\mathrm{\ref{fn:Out-of-range.}}$ & 85 & \multicolumn{2}{c}{-} & \multicolumn{2}{c}{-}\tabularnewline
\multirow{2}{*}{109} & \multirow{2}{*}{-} & \multicolumn{2}{c}{108 } & 3&25\tabularnewline
 &  & \multicolumn{2}{c}{111 } & 3&32\tabularnewline
120\footnote{In the metallic state at room $T$ (our data).} & - & \multicolumn{2}{c}{125 } & 3&76\tabularnewline
- & 139 & \multicolumn{2}{c}{-} & \multicolumn{2}{c}{-}\tabularnewline
145 & - & \multicolumn{2}{c}{-} & \multicolumn{2}{c}{-}\tabularnewline
172 & - & \multicolumn{2}{c}{173 } & 5&2\footnote{Barely observable by averaging several low-$T$ spectra.\label{fn:Barely-observable-by}}\tabularnewline
203 & - & \multicolumn{2}{c}{200 } & 5&69$^\mathrm{\ref{fn:Barely-observable-by}}$\tabularnewline
\end{tabular}\caption{Comparison of the observed low-$T$ Raman-/coherent-mode frequencies
in different experiments.}
\label{tab:freq}
\end{table}

The high-$T$ space group of CuIr$_{2}$S$_{4}$ is $\mathrm{F}d\overline{3}m$
leading to five distinct Raman lines corresponding to the $1A_{g}+1E_{g}+3F_{2g}$
irreducible representations.\citep{zhangLing2010} The Raman tensor
of the $A_{g}$ mode is diagonal and isotropic\citep{KroumovaAroyo2003}
so the scattering in the back scattering geometry is allowed for parallel
excitation-scattering polarizations and suppressed for the perpendicular
polarizations for any light propagating direction. 

The nonzero Raman tensor components for the degenerate $E_{g}$ representations
are (i) $\chi_{xx}^{E_{g}}(1)=\chi_{yy}^{E_{g}}(1)=b$, $\chi_{zz}^{E_{g}}(1)=-2b$
and (ii) $\chi_{xx}^{E_{g}}(2)=-\chi_{yy}^{E_{g}}(2)=-\sqrt{3}b$.
The scattering \footnote{Simultaneously for both degenerate modes.}
is suppressed only for the crossed polarizations parallel to the <100>
directions.

The nonzero Raman tensor components for the triply-degenerate $F_{2g}$
representations are: $\chi_{ij}^{F_{2g}}(k)=\chi_{ji}^{F_{2g}}(k)$
where $ij\equiv xy$, $xz$ and $yz$ for $k=1$, 2 and 3, respectively.
The scattering is suppressed only for the crossed polarizations parallel
to the <110> directions.

In sample S1 only the $A_{g}$ mode is suppressed in the crossed polarization
due to the low symmetry surface orientation. On the other hand, in
sample S2 systematic extinctions consistent with the back-scattering
along the {[}001{]} direction were observed for different polarization
configurations implying that the orientation of the S2 cleaved surface
corresponds to a  <001> plane.

The observed polarization dependence of the additional weak line at
120 cm$^{-1}$ is consistent with the $F_{2g}$ representation in
both samples and is therefore attributed to the previously unobserved\citep{zhangLing2010}
$F_{2g}$(3) modes.

The coherent-mode frequencies and the number of the observed coherent
modes are different than reported previously\citep{MatsubaraEbihara2009}
(see Table \ref{tab:freq}). Sample S2 has identical orientation\footnote{Due to the low point symmetry of the low-$T$ insulating phase it
is also very unlikely that the difference could be due to the different
orientations of the measured surfaces and the correspondingly different
selection rules.} of the measured surface as reported in Ref. {[}\onlinecite{MatsubaraEbihara2009}{]}
and the excitation and probe photon energies are also the same. Moreover,
the modes reported in Ref. {[}\onlinecite{MatsubaraEbihara2009}{]}
are insensitive to the IM transition. It is therefore very likely
that the transient response reported in Ref. {[}\onlinecite{MatsubaraEbihara2009}{]}
is not intrinsic.

A better correspondence is found to the low-$T$ Raman mode frequencies\citep{zhangLing2010}.
In the overlap region\footnote{The published\citep{zhangLing2010} Raman spectra start at 100 cm$^{-1}$
and the present experiment time resolution limits the maximum coherent-mode
frequency below the cutoff of $\sim300$ cm$^{-1}$.} we observe a correspondence of the coherent mode doublet at 108/111
cm$^{-1}$ to the broad Raman line at 109 cm$^{-1}$. We also observe
weak signs of modes\footnote{The modes are so weak that a reliable identification is possible only
by averaging all transients measured below $T_{\mathrm{IM}}$.} (not shown in Fig. \ref{fig:coherent}) at the 172-cm$^{-1}$ and
203-cm$^{-1}$ Raman mode positions. 

On the other hand, the coherent mode at 125 cm$^{-1}$ is absent in
the low-$T$ Raman spectra\citep{zhangLing2010} and the low-$T$
Raman mode\citep{zhangLing2010} at 145 cm$^{-1}$ is absent in the
coherent response. However, the 120-cm$^{-1}$ room-$T$ frequency
of the newly observed weak $F_{2g}(3)$ mode is close to the 125-cm$^{-1}$
coherent mode frequency so the mode could be linked to the $F_{2g}(3)$
mode that hardens\footnote{The coherent mode frequency just below $T_{\mathrm{IM}}$ is 122.7
cm$^{-1}$.} and changes the symmetry to $A_{g}$ upon entering the low-$T$ insulating
state. In the high-$T$ state it is absent from the coherent response
since it cannot be excited by the displacive mechanism due to the
$F_{2g}$ symmetry. According to the low-$T$ symmetry the mode should
also split to three $A_{g}$ modes\footnote{For the approximate\citep{radaelliHoribe2002} $I4_{1}/amd$ ($D_{4h}$)
group the splitting would be twofold into $B_{1g}+E_{g}$ modes. }, but this is not observed in the experiment. 

Since there is no selection-rules-forbidden\footnote{Due to the low point-group symmetry (C$\mathrm{_{i}}$) of the low-$T$
phase all Raman active modes correspond to the $A_{g}$ representation. } modes at low-$T$ in both experiments we attribute the absence of
the 145-cm$^{-1}$ mode in the coherent response to different Raman
resonance conditions.

The low-$T$ lattice-structure\citep{radaelliHoribe2002} primitive
cell is large and has low symmetry (P$\bar{1}$) so 84 Ag modes are
Raman allowed.\citep{KroumovaAroyo2003} Looking at the low frequency
range the heaviest Ir ions contribute 24 and Cu ions further 12 $A_{g}$
modes to the mechanical representation while in the high-$T$ phase
Ir ions do not contribute to the Raman allowed modes at all.

In experiments only 8 modes are observed below $\sim200$ cm$^{-1}$.
This is consistent with proximity\citep{furubayashiMatsumoto1994,radaelliHoribe2002}
of the structure to the $I4_{1}/amd$ ($D_{4h}$) symmetry where the
Ir ions are not Raman active and the Cu ions contribute to the one
$B_{1g}$ and one $E_{g}$ symmetry mode only, which corresponds to
the 120-cm$^{-1}$ $F_{2g}$(3) mode of the high-$T$ phase.

Among the 8 observed modes the 125-cm$^{-1}$ (3.76-THz) related $F_{2g}$(3)
mode does not contain Ir-ions displacements in the high-$T$ phase
so it is expected to remain Cu-ions displacements dominated in the
low-$T$ phase as well. The two lowest frequency modes at 62 cm$^{-1}$
(1.86 Thz) and 78 cm$^{-1}$ (2.34 Thz) are therefore expected to
be the most closely related to the vibrations of the dimerized Ir$^{+4}$
ions that shift the most at the IM phase transitions. 

Turning to the non-oscillating relaxation dynamics it is unexpected
that there is no clear relaxation component due to the recombination
of the photoexcited quasiparticle across the gap in the low-$T$ state.
The observed 15-ps exponential relaxation is not correlated with the
gap opening since it clearly appears only well below the $T_{\mathrm{IM}}$.
\footnote{From the present data it is not clear whether the 15-ps exponential
relaxation is the photoexcited quasiparticle related since it can
also correspond to the incoherent-phonons induced band renormalizations.} The experimentally reported gap\citep{wangCao2004optical} of 150
meV (1200 cm$^{-1}$) significantly exceeds the maximum phonon frequency
of $<\sim500$ cm$^{-1}$. A relaxation slower than our experimental
time window is therefore expected due to the recombination since single
phonon processes are forbidden. The absence of such component therefore
suggests that the optical transitions at 1.55-eV do not involve states
at the gap edges.

\section{Summary and conclusions}

We performed a room-$T$ Raman and systematic $T$-dependent investigation
of the ultrafast transient reflectivity in CuIr$_{2}$S$_{4}$.

In the Raman spectra we observe an additional previously undetected
$F_{2g}$ mode at 120 cm$^{-1}$. 

The transient reflectivity in the low-$T$ insulating phase at 1.55-eV
photon energy is dominated by the broken-lattice-symmetry-induced
coherent lattice response that abruptly vanishes at the insulator-metal
transition. The frequencies of the coherent modes are different from
previously reported transient reflectivity study\citep{MatsubaraEbihara2009},
but consistent with the published low-$T$ Raman spectra\citep{zhangLing2010}.
The coherent modes show a rather standard anharmonic $T$-dependence
in the full insulating phase $T$-range.
\begin{acknowledgments}
The authors acknowledge the financial support of Slovenian Research
Agency (research core funding No-P1-0040 and ) for financial support.
We would also like to thank V. Nasretdinova and E. Goreshnik for the
help at the sample characterization.
\end{acknowledgments}

\section{Appendix}

\subsection{DECP model fits}

To analyze the signal we fit the data using the DECP model\citep{zeiger1992theory}
where the transient reflectivity is given by 
\begin{gather}
\frac{\Delta R}{R}=(A\mathrm{_{displ}}+\sum A_{\mathrm{O}i})\int_{0}^{\infty}G(t-u)e^{-u/\tau_{\mathrm{displ}}}du\nonumber \\
-\sum A_{\mathrm{O}i}\int_{0}^{\infty}G(t-u)e^{-\gamma_{i}u}[\cos(\Omega_{i}u)\nonumber \\
-\beta_{i}\sin(\Omega_{i}u)]du\nonumber \\
+\sum A_{\mathrm{e}j}\int_{0}^{\infty}G(t-u)e^{-u/\tau_{j}}du,\label{eq:fitfunc}
\end{gather}
where $\beta_{i}=(1/\tau_{\mathrm{displ}}-\gamma_{i})/\Omega_{i}$
and $G(t)=\sqrt{\nicefrac{2}{\pi}}\tau_{\mathrm{p}}\exp(-2t^{2}/\tau_{\mathrm{p}}^{2})$
with $\tau_{\mathrm{p}}$ being the effective pump-probe pulse cross-correlation
width. In the DECP model the coherent modes are driven with an exponentially
relaxing displacive mode with the relaxation time $\tau_{\mathrm{displ}}$
and the amplitude, $A\mathrm{_{displ}}$\footnote{$A\mathrm{_{displ}}$ represents the coupling of the displacive mode
to the optical reflectivity while the couplings to different oscillatory
modes are implicitly included in parameters $A_{\mathrm{O}i}$.}. $A_{\mathrm{O}i}$, $\Omega_{i}$, $\gamma_{i}$ are the oscillating
modes amplitudes, frequencies and damping factors, respectively, while
$A_{\mathrm{e}j}$ and $\tau_{j}$ are the amplitudes and relaxation
times of additional exponentially relaxing modes.

\bibliographystyle{apsrev4-1}
\bibliography{biblio}

\end{document}